\documentclass[onecolumn,preprintnumbers,amsmath,amssymb]{revtex4}

\usepackage{graphicx}
\usepackage{dcolumn}
\usepackage{bm}

\begin{document}

\title{Second harmonic generation in gallium phosphide photonic crystal nanocavities with ultralow continuous wave pump power}

\author{Kelley Rivoire$^{1*}$, Ziliang Lin$^1$, Fariba Hatami$^2$,  W. Ted Masselink$^2$, and Jelena Vu\v{c}kovi\'{c}$^1$}

\address{$^1$E. L. Ginzton Laboratory, Stanford University, Stanford, CA 94305-4085\\ $^2$Department of Physics, Humboldt University, D-10115, Berlin, Germany}

\email{krivoire@stanford.edu} 



\begin{abstract}
 We demonstrate second harmonic generation in photonic crystal nanocavities fabricated in the semiconductor gallium phosphide. We observe second harmonic radiation at 750 nm with input powers of only nanowatts coupled to the cavity and conversion efficiency $P_{\rm out}/P_{\rm in, coupled}^2 = 430\%/{\rm W}$. The large electronic band gap of GaP minimizes absorption loss, allowing efficient conversion. Our results are promising for integrated, low-power light sources and on-chip reduction of input power in other nonlinear processes.
\end{abstract}
\maketitle

\section{Introduction}
III-V
semiconductors such as GaAs and GaP are considered promising candidates for nonlinear optical devices\cite{levi, kuo, vodopyanov} because of their
large second order nonlinearity\cite{shoji}, transparency over a wide wavelength range (870 nm-17 $\mu$m for GaAs and 550 nm-11 $\mu$m for GaP), and ease of integration with semiconductor processing. The cubic symmetry of the zincblende lattice of III-V semiconductors, however, does not exhibit
birefringence, so achieving phase matching of the two different frequencies typically requires employing quasi-phase matching techniques or an additional birefringent material \cite{eyres, scaccabarozzi}.

In addition to the challenges posed by phase matching, nonlinear optical devices are also constrained by the bulky, macroscopic resonant cavities typically used to enhance conversion efficiency and reduce the required input power\cite{byer}. High quality factor microcavities have the potential to achieve similar conversion efficiencies with a vastly reduced size, and could be integrated with nanophotonic technology. In these microcavities, the phase matching condition is satisfied by the spatial overlap between the fundamental and second harmonic field patterns\cite{johnson, orenstein, liscidini:1883}. Experimentally, this resonant enhancement has been demonstrated in silica microdisks\cite{vahala}, where green third harmonic radiation was
observable with hundreds of microwatts incident continuous wave IR power. Enhanced second harmonic generation in photonic crystal cavities\cite{mccutcheon}and third harmonic generation in photonic crystal waveguides\cite{corcoran} waveguides have also been demonstrated. However, these experiments suffered from poor conversion efficiency, in part because the second harmonic was above the band gap of the employed semiconductor and therefore strongly absorbed in it (where efficiency is defined as the ratio of the output second harmonic to the coupled input power at 1550 nm, $P_{\rm out}/P_{\rm in, coupled} = 5 \times 10^{-10}$ for peak $P_{\rm in}=1W$ \cite{corcoran}, $P_{\rm out}/P_{\rm in} = 10^{-13}$ for CW $P_{\rm in}=300 \mu {\rm W}$ \cite{mccutcheon}). In this paper, we show that photonic crystal nanocavities resonant with the pump wavelength can be used to generate second harmonic radiation with input power orders of magnitude smaller than previously demonstrated.

\section{Cavity design}
The experiment is described in Figure 1. Our resonator is a modified linear three hole defect photonic crystal cavity\cite{noda}. A scanning electron
microscope (SEM) image is shown in Figure 1b.  A tunable infrared laser (Agilent 81680A) with wavelength range around 1500 nm at normal incidence is spatially aligned to the cavity location and spectrally aligned to the cavity resonance; the laser polarization is also aligned to match that of the cavity mode. In this configuration, enhanced second harmonic radiation is generated; this radiation is either analyzed by a spectrometer, measured by a femtowatt photodetector, or imaged onto a camera. A second harmonic spectrum with 8 nW power coupled to the cavity (160 nW incident on the sample) is shown in Figure 1c. The electronic band gap of GaP prevents both absorption of the harmonic radiation at 750 nm and two photon absorption at the fundamental wavelength 1500 nm. The samples were grown by gas-source molecular beam epitaxy on a (100)-oriented GaP wafer. A 160 nm thick GaP membrane was grown on the top of a 1 $\mu$m thick sacrificial AlGaP layer. Structures were fabricated with e-beam lithography and etching, as described in \cite{rivoire}. The photonic crystal cavities are three hole linear defects\cite{noda} with lattice constant $a=500-560 nm$, hole radius $r/a\approx0.2-0.25$, and slab thickness $d/a\approx0.3$. We use a perturbation design for our photonic crystal cavities \cite{toishi} to increase the signal collected when characterizing the fundamental resonance.

To efficiently radiate power at the second harmonic, the field patterns at the fundamental and second harmonic frequency must have constructive overlap. In our experiment, the fundamental frequency is set to that of the cavity mode, which dictates the field pattern at this frequency. The second harmonic frequency is matched to a higher order photonic band edge mode, which determines the second harmonic field pattern. We simulate the field patterns inside the structure at the fundamental and second harmonic wavelengths using 3D finite difference time domain methods. The in-plane electric field profile in the center of the slab for the fundamental transverse electric-like (TE-like) cavity mode is shown in Figure 2a. In the far-field, this cavity radiates primarily with $\hat{y'}$ polarization, and therefore incident light with this polarization can couple into the cavity. GaP has a noncentrosymmetric cubic crystal structure; the only non-zero elements of the bulk $\chi^{(2)}_{ijk}$ have $i\neq j \neq k$. Since the input signal in our experiment couples to the TE-like photonic crystal cavity mode (with dominant $E'_x$ and $E'_y$ in-plane field components defined in Fig. 2a, 2b), the output second harmonic signal will then be primarily $\hat{z}$-polarized, i.e., transverse magnetic-like (TM-like) mode. The bulk second harmonic nonlinear polarization generated takes the form $P^{(2)}_z =2\epsilon_0 d_{14} E_x E_y$ where $x, y$ are the crystal axes, $E_x$, $E_y$ are the corresponding components of electric field of the coupled input signal (pump), $\epsilon_0$ is the permittivity of free space, and $d_{14}$ is the second-order nonlinear coefficient in contracted notation. The photonic band diagram for TM-like modes is shown in Figure 2c. Modes near the $\Gamma$ point ($k_x= k_y$ = 0) at the second harmonic wavelength can efficiently radiate into the numerical aperture of our objective lens (indicated by black solid lines). The mode with frequency closest to the second harmonic is indicated. The field patterns for this doubly degenerate mode (Figure 2d) indicate that spatial overlap between the fundamental and second harmonic fields is best (i.e. with the $E_z$ field pattern that best matches $P^{(2)}_z$ described by the previous expression, maximizing $|\int dV E_z P_z|$) for cavities oriented at $\theta = 0^{\circ}$, in agreement
with our experimental observations for cavities with different $\theta$. The fundamental and second harmonic field patterns have imperfect spatial overlap; future optimization of this overlap, or introduction of a TM cavity mode at the second harmonic frequency\cite{yinan}, would improve conversion efficiency.

\section{Experimental characterization}
We first characterize the fundamental resonance of the cavities by probing them from vertical incidence using cross-polarized reflectivity with a broadband lamp\cite{rivoire} and a tunable laser. The cross-polarized configuration is used to obtain sufficient signal-to-noise to observe the resonance above the reflected background uncoupled to the cavity. We select cavities with fundamental TE-like modes at 1480 nm-1560 nm, within the
range of our tunable laser. We estimate the coupling efficiency into the cavity to be approximately 5\% by measuring the reflected and incident light power, in agreement with simulations for this type of cavity\cite{toishi}. (For higher input coupling efficiency, the cavity could be integrated with photonic crystal waveguide couplers\cite{faraon}.)  A typical reflectivity spectrum measured with a tungsten halogen white light source is shown in  Figure 3a; the measured quality factor is 5600.  Finite difference time domain (FDTD) simulations indicate that the Q-factor in this structure is limited to 15000  because of the thin slab. The spectral profile we observe is governed by the Lorentzian density of states of the cavity   $\rho_c\left(\omega\right)=\frac{1}{\pi}\frac{\omega/2Q}{{(\omega/2Q)^2+\left(\omega-\omega_c\right)}^2}$ where $\omega_c$ and $Q$ are the cavity frequency and quality factor respectively.  Once we locate the fundamental resonance of the cavity,  we use the continuous wave tunable infrared laser to generate second harmonic radiation, as depicted in Figure 1.  A typical spectrum of the second harmonic signal as we scan the laser through the cavity resonance is shown in Figure 3b. The  cavity enhances the coupled input power by a factor proportional to $Q^2$ on resonance; away from resonance, this enhancement drops off like a Lorentzian  squared. On resonance, the total second harmonic power $P_{\rm out}$ emitted depends on the square of the input power coupled to the cavity $P^2_{\rm in, coupled}$, the square of the quality factor $Q$ of the cavity, and the overlap between fundamental and second harmonic mode field profiles,  $P_{\rm out}\propto P^2_{\rm in, coupled} Q^2 |\int d_{14}^{(2)} E_{x,\omega} E_{y,\omega} E^*_{z,2\omega} dV|^2$.  From the second harmonic spectrum, we can extract the quality factor of the cavity  mode equal to 6000, in good agreement with the Q observed in the reflectivity measurements at the fundamental wavelength (1500 nm).

To measure the absolute second harmonic power radiated, we send the second harmonic signal to a femtowatt photodetector. Figure 4 shows the measured output power as a function of the incident power coupled into the cavity, assuming an input coupling to the cavity of 5\%. The data show a good fit to a quadratic dependence on input power. We measure a normalized conversion efficiency $P_{\rm out}/P_{\rm in, coupled}^2$ of 430\%/W, or $P_{\rm out}/P_{\rm in, coupled} = 5 \times 10^{-5}$ for $11  \mu {\rm W}$ coupled input power ($P_{\rm out}/P_{\rm in} = 2 \times 10^{-6}$ for $220  \mu{\rm W}$ power through the objective lens). The measured value is an underestimate of the second harmonic power generated inside the cavity, as we only measure radiation that is directed vertically and collected with our objective lens.

We also investigate the dependence of the second harmonic signal on the polarization of the incident light (Fig. 5a). To couple power into the cavity and generate resonantly enhanced second harmonic radiation, the incident light must have polarization aligned to the polarization the cavity
radiates. For a cavity oriented at $45^\circ$ relative to the crystal axes, we see maximum second harmonic signal when the input polarization is aligned to the cavity mode polarization, and a reduction from this maximum by a factor of $\cos^4\left(\theta\right)$ as the incidence angle is rotated away from the cavity, as expected for a quadratic process. We observe a similar angular dependence for cavities oriented at other angles to crystal axes. Figures 5b and 5c show images of second harmonic radiation recorded on a camera (DVC 710M, 10s integration time).

\section{Conclusions}
In summary, we demonstrate efficient second harmonic generation in the visible from photonic crystal cavities with incident powers well below $1  \mu{\rm W}$ as a result of resonant recirculation of the pump (1500 nm) light in the GaP nanocavity. By using a GaP membrane for our structures, we minimize absorption losses at the output wavelength of 750 nm. Our structures could serve as on-chip, low-power sources compatible with semiconductor fabrication processing. These results also indicate the potential of microcavities for significantly reducing the required input powers for other experiments in nonlinear optics, such as sum/difference frequency generation and parametric downconversion.

We thank Professors Marty Fejer and Steve Harris at Stanford University for helpful comments on a draft of the manuscript. Financial support was provided by the National Science Foundation (NSF Grant DMR-0757112). K.R. and Z.L. are supported by National Science Foundation Graduate Research Fellowships and Stanford Graduate Fellowships. The work was performed in part at the Stanford Nanofabrication Facility of NNIN.

 \begin{figure*}[htb]
\includegraphics{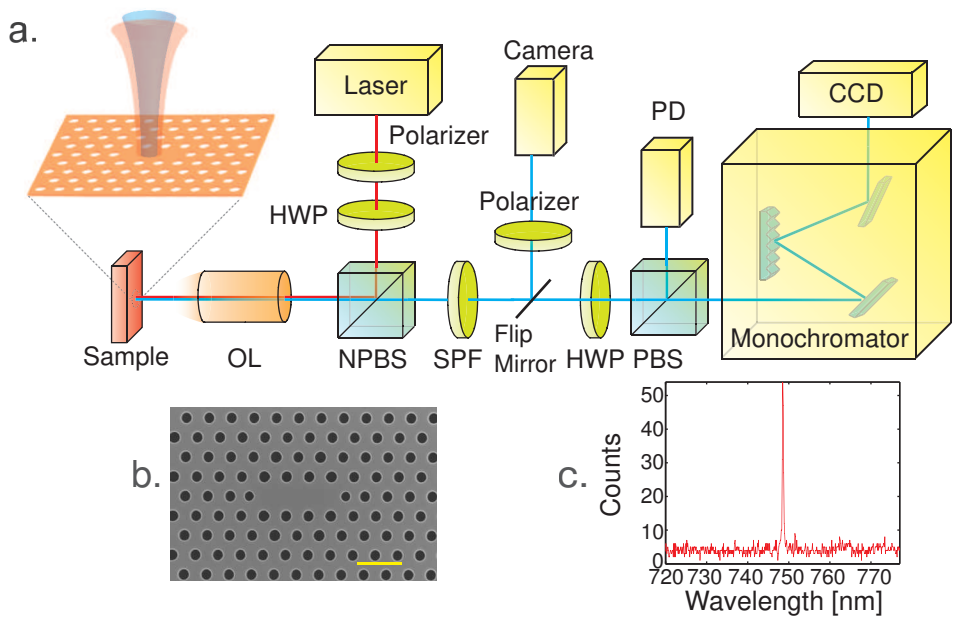}
\caption{\label{fig:pc2}
(a) Confocal microscope-based setup for second harmonic generation.
HWP: half wave plate, NPBS: nonpolarizing beamsplitter, OL: objective lens, PBS: polarizing beamsplitter, SPF: short pass filter, PD: photodiode. The incident light traces the red line into the cavity sample. The second harmonic light follows the blue line into the spectrometer, photodiode, or camera. The
polarization of the incident light is controlled by the polarizer
and HWP; the polarization of the second harmonic
radiation is measured using HWP and PBS. (b) SEM image of a fabricated structure. Scale
bar indicates 1 $\mu$m.
 (c) Spectrum of generated second harmonic light with 8 nW power at 1500 nm coupled to the cavity (160 nW incident).}
\end{figure*}
 \begin{figure*}[htb]
\includegraphics{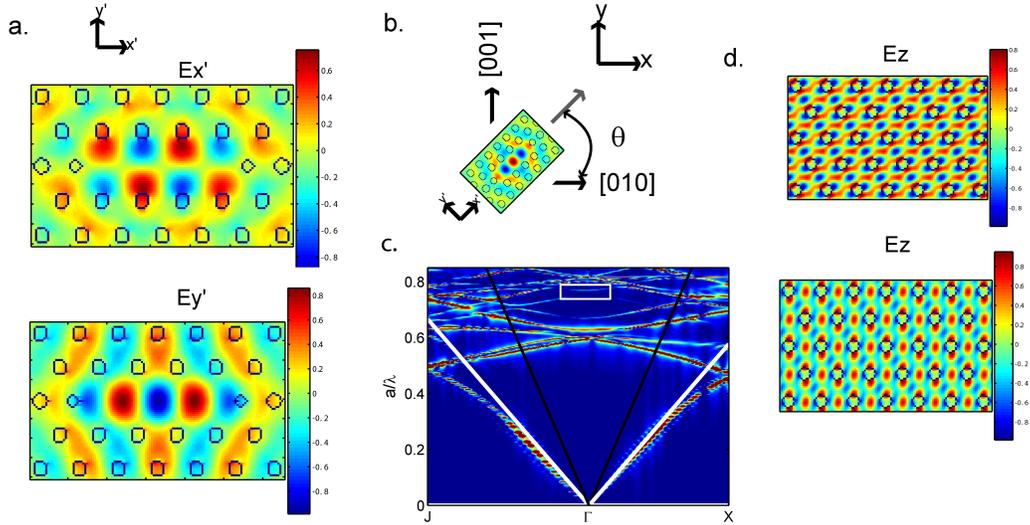}
\caption{\label{fig:pc22}
(a) Finite difference time domain (FDTD) simulation
of electric field inside the cavity for the fundamental TE-like cavity
resonance in the center of the slab. Cavity field axes are $E'_x$ and $E'_y$ (b) Illustration of orientation of cavity relative to crystal axes. Cavities axes $E'_x$, $E'_y$ are rotated from crystal axes $E_x$, $E_y$ by an angle $\theta$. Fields along crystal axes are determined by projection.
(c) FDTD simulation of TM-like photonic bands for same triangular lattice photonic crystal. Red indicates band positions. White solid lines indicate light line; black solid lines indicate numerical aperture of lens. White box indicates mode at second harmonic frequency. $a$: lattice constant of photonic crystal. (d) $E_z$ field patterns of degenerate TM-like mode at second harmonic frequency (white box in Fig. 2c.)
}
\end{figure*}

\begin{figure*}[htb]
\includegraphics{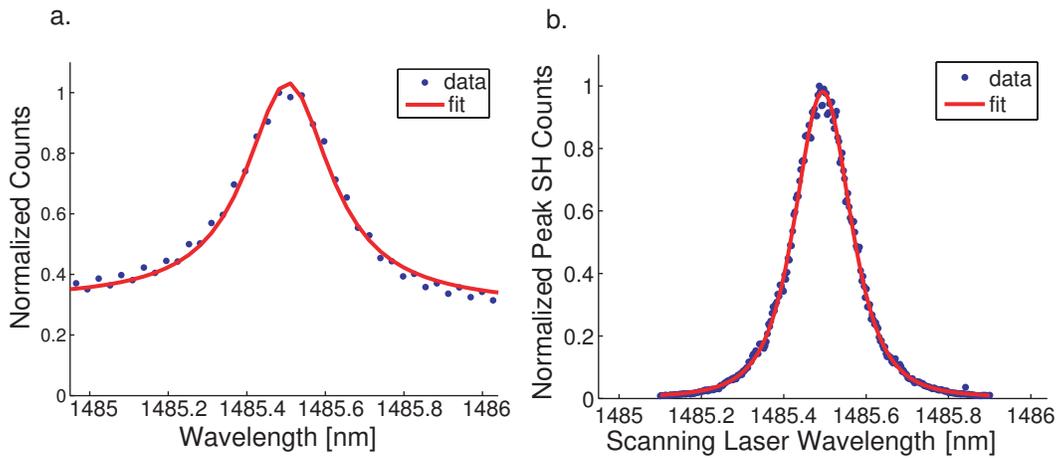}
\caption{\label{fig:pc2}
(a) Spectrum of fundamental resonance probed in cross-polarized reflectivity with a broadband source.
Lorentzian fit gives a quality factor of 5600 (b) Spectrum at second harmonic as exciting laser frequency is tuned across the cavity resonance.
Solid line shows fit to Lorentzian squared with cavity quality factor of 6000.}
\end{figure*}

\begin{figure}[htb]
\includegraphics[width=10cm]{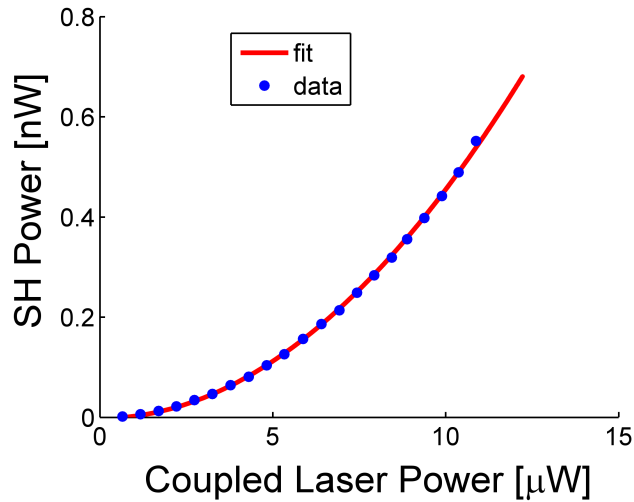}
\caption{\label{fig:pc}
(a) Second harmonic power as a function of
fundamental wavelength power coupled into the cavity. We estimate
coupling efficiency into the cavity to be 5\%. Solid line indicates
quadratic fit. Output power
measurements are corrected for measured losses from optics, but do
not include corrections for collection efficiency into the
objective lens. }
\end{figure}

\begin{figure*}[htb]
\includegraphics{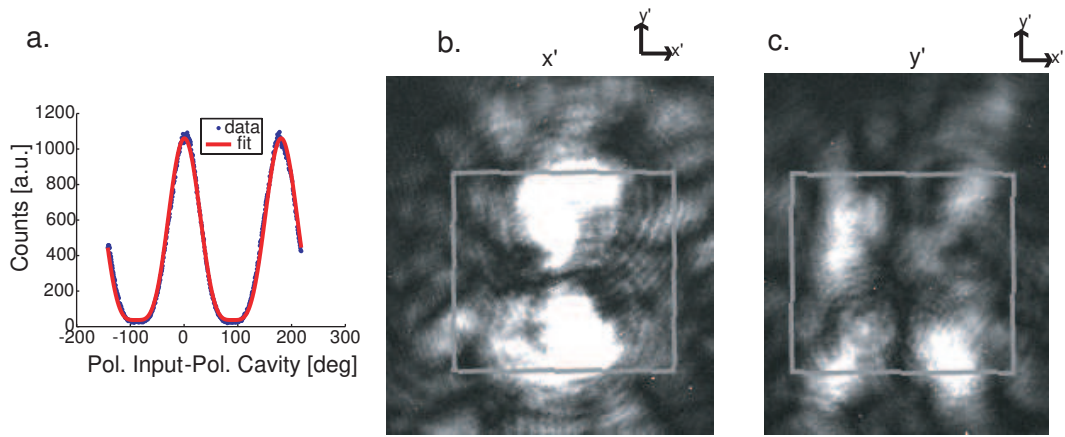}
\caption{\label{fig:pc}
(a) Dependence of second harmonic power on
incident light polarization. The horizontal axis corresponds to the angle between the input polarization and the polarization of the cavity mode. Solid line shows fit to $\cos^4(\theta)$. (b) and (c) Second harmonic radiation imaged on a camera with polarizer oriented in $x'$ (b) and $y'$ (c) direction (the orientation of cavity relative to the axes is shown in Fig. 2b). The gray box indicates the approximate area of the photonic crystal structure.
 }
\end{figure*}


\end{document}